\documentclass[a4paper,fleqn]{cas-dc}

\usepackage{soul,color}
\usepackage[numbers, sort&compress]{natbib}
\usepackage{lipsum}
\usepackage{booktabs}
\usepackage{comment}
\usepackage[normalem]{ulem}

\def\tsc#1{\csdef{#1}{\textsc{\lowercase{#1}}\xspace}}
\tsc{WGM}
\tsc{QE}
\tsc{EP}
\tsc{PMS}
\tsc{BEC}
\tsc{DE}

\begin{document}
\let\WriteBookmarks\relax
\def\floatpagepagefraction{.8}
\def\textpagefraction{.1}
\shorttitle{Attractors HF}
\shortauthors{S. Chen et~al.}

\title [mode = title]{Non-equilibrium Dynamical Attractors and Thermalisation of Charm Quarks in Nuclear Collisions at the LHC Energy} 



\author[2]{Shile Chen}[orcid=0000-0002-3874-5564]
\ead{shchen@lns.infn.it}

\author[2]{Vincenzo Nugara}[orcid=0009-0006-1939-8663]
\ead{vincenzo.nugara@phd.unict.it}

\author[1,2]{Maria Lucia Sambataro}[orcid=0009-0007-9018-661X]
\ead{sambataro@lns.infn.it}

\author[1,2]{Salvatore Plumari} [orcid=0000-0002-3101-8196]
\ead{salvatore.plumari@dfa.unict.it}

\author[1,2]{Vincenzo Greco}[orcid=0000-0002-4088-0810]

\address[1]{Dipartimento di Fisica e Astronomia "E. Majorana", Università degli Studi di Catania, Via S. Sofia 64, 1-95125 Catania, Italy}
\address[2]{Laboratori Nazionali del Sud, INFN-LNS, Via S. Sofia 62, I-95123 Catania, Italy}

\begin{abstract}
We study the non-equilibrium dynamics, thermalisation and attractor behaviour of charm quarks in a longitudinally expanding Quark–Gluon Plasma within the Relativistic Boltzmann Transport approach in 1+1D Bjorken expansion.
Considering both a strong AdS/CFT coupling scenario with constant $2\pi T D_s=1$ and a temperature-dependent diffusion coefficient 
$D_s^\text{lQCD}(T)$ from the recent unquenched lattice QCD data, we analyse the evolution of effective temperature, momentum moments and distribution functions for different initial conditions, including FONLL and EPOS4HQ spectra.
We find that charm quarks exhibit dynamical attractors; however, the temperature dependence of $D_s^\text{lQCD}(T)$ leads to significantly longer relaxation times compared to the strong coupling limit. While dynamical attractors occur within $
\sim 1-1.5 \rm \,fm$ for $2\pi T D_s=1$, they are delayed to 
$\sim 5 \rm \,fm$ for $D_s^\text{lQCD}(T)$, becoming comparable to 
the lifetime of the Quark-Gluon Plasma phase in ultra-relativistic collisions. This indicates that charm quarks may not fully thermalise, especially in small systems such as peripheral or light-ion collisions.
We further show that, for $D_s^\text{lQCD}(T)$, the deviation from equilibrium becomes as large as
$\delta f_{HQ}/f_{eq} \sim p_T^\beta \sim \mathcal{O}(1)$ already at $p_T\simeq 3\rm\, GeV$, rising with $\beta \sim 4.5$, thus questioning the applicability of viscous hydrodynamics to charm dynamics.
\end{abstract}


\maketitle

\section{Introduction}
The theoretical studies on Hot QCD matter and the phenomenological analysis of the abundant experimental observables of ultra-Relativistic Heavy-Ion Collisions (uRHICs) at RHIC and LHC indicate the production of a deconfined phase in which the degrees of freedom are quarks and gluons, the Quark-Gluon Plasma (QGP). Characterising the properties of this Hot QCD matter and accessing the initial state of the collision starting from final observables is a challenging task, and several approaches have been proposed in the last two decades. A powerful tool of investigation are the Heavy Quarks (HQs), more specifically charm and bottom ones, due to the extremely short lifetime of the top. Since their masses are larger than the typical maximum temperature of the medium ($T_0\sim 0.5-0.6$ GeV at top LHC energies), they are mainly created by initial hard scattering processes, with their numbers being conserved through the whole evolution of the medium. Moreover, their dynamics in the QGP is usually modelled as a Brownian motion due to their large masses.
This framework has been extensively used to calculate key observables 
\cite{Dong:2019unq,He:2022ywp,Scardina:2017ipo,vanHees:2005wb,vanHees:2007me,Gossiaux:2008jv,Das:2009vy,Alberico:2011zy,Uphoff:2012gb,Lang:2012nqy,Song:2015sfa,Song:2015ykw,Das:2013kea,Cao:2015hia,Das:2015ana,Cao:2017hhk,Das:2017dsh,Jamal:2020fxo,Ruggieri:2018rzi,Sun:2019fud,Cao:2018ews,Rapp:2018qla,Sambataro:2023tlv,Sambataro:2022sns,Plumari:2019hzp}, such as the nuclear suppression factor $(R_{AA})$ which describes how the HQ spectra are modified in $AA$ collisions with respect to the $pp$ ones and  the elliptic flow $v_2=\langle \cos(2\phi_p) \rangle$, a measure of the anisotropy in the angular distribution of heavy flavour hadrons. Both experimental findings and phenomenological investigations have been collecting hints of a partial thermalisation for charm quarks. 
Indeed, the observation of positive anisotropic flows for open charm hadrons and charmonia, similar to those observed for the light sector \cite{CMS:2017vhp, ALICE:2020iug, ALICE:2020pvw}, has even stimulated investigations about the possible applicability of hydrodynamics to study their evolution \cite{Capellino:2022nvf, Capellino:2023cxe, Facen:2025phz}. Moreover, recent lattice QCD (lQCD) calculations with dynamical fermions indicate a significant low value of the spatial diffusion coefficient $2\pi T D_s\simeq 1$ at $T\simeq T_c$ for charm quarks \cite{Altenkort:2023oms,Altenkort:2023eav,HotQCD:2025fbd}, suggesting at least for $p \rightarrow 0$ a quite short thermalisation time, $\tau_{eq}\sim 1-2 \, \rm fm$, compatible with the AdS/CFT predictions.\\
In this perspective, charm quarks dynamics enters the broader field studying the regime of applicability of relativistic hydrodynamics, that
unexpectedly succeeded in describing observables for heavy-ion collisions and could be able to describe much smaller systems, such as $pp$ or $pA$ \cite{Weller:2017tsr}, whose lifetime and size should prevent them from reaching even a partial thermalisation. 
In order to study why and how the so-called ``hydrodynamisation'' process takes place, different approaches have been followed: one of the most fruitful has been the study of dynamical attractors. In systems with an initial strong longitudinal expansion, distinct initial conditions evolve towards a universal behaviour well before reaching partial equilibration, suggesting a decay of degrees of freedom towards a few macroscopic variables which could be interpreted as a hint of the applicability of hydrodynamics. Dynamical attractors have been found in distinct frameworks, including hydrodynamics, kinetic theory, classical Yang--Mills dynamics and AdS/CFT 
\cite{Heller:2015dha, Strickland:2018ayk,
Almaalol:2020rnu, Noronha:2021syv,
Blaizot:2017ucy, Soloviev:2021lhs, Cartwright:2022hlg, Strickland:2017kux,
Kurkela:2019set, Spalinski:2018mqg,
Denicol:2019lio, Behtash:2017wqg, Chattopadhyay:2019jqj, Giacalone:2019ldn, Mazeliauskas:2018yef, Brewer:2022vkq, Berges:2013fga, Chen:2025qao, Jankowski:2023fdz}. Most of the research work has been carried out for a conformal one-component 0+1D Bjorken system, while some investigations have been performed for non-conformal systems \cite{Alqahtani:2022dfm, Alalawi:2022pmg, Spalinski:2025ngd}, 1+1D and 3+1D scenarios \cite{Ambrus:2021sjg, Chen:2024pez,  Nugara:2023eku, Nugara:2024net} and mixtures \cite{Frasca:2024ege}.\\
In this Letter, we present a first study of the evolution of charm quarks in a medium in the simplified scenario of Bjorken flow in 1+1D, addressing the issue of the existence of dynamical attractors, universal behaviour and thermalisation. We aim to identify the time scales within which thermalisation is reached in various regions of the phase space for different coupling regimes; moreover, we quantify the deviation from equilibrium of the charm distribution function, which sensitively affects predictions on experimental observables \cite{Facen:2025phz}. 


\section{Transport evolution of charm quark in the QGP}

The results shown in this paper have been obtained using the Relativistic Boltzmann Transport (RBT) code developed in recent years to perform studies of the dynamics of relativistic heavy-ion collisions at both RHIC and LHC energies \cite{Plumari:2012ep,Ruggieri:2013ova,Scardina:2014gxa,Plumari:2015cfa,Scardina:2017ipo,Plumari:2019gwq,Sun:2019gxg, Nugara:2023eku, Nugara:2024net}. 
We are employing a bulk with massive light quarks and gluons with mass $m_i=0.5$ GeV resembling, in the explored range of temperature, typical mass value in the Quasi-Particle Model \cite{Sambataro:2024mkr}, which allows us to describe an Equation of State in agreement with lQCD \cite{Borsanyi:2020fev, HotQCD:2025fbd}. We consider here only charm quarks, fixing the mass of the heavy flavour sector $m_{HQ}=1.3$ GeV.\\
In our approach, the space-time evolution of the distribution functions of bulk matter, made up of gluons ($g$) and light quarks ($q$), as well as of charm quarks ($HQ$), is described by means of the coupled Relativistic Boltzmann equations given by:
	\begin{eqnarray}
		& & p^{\mu}_i \partial_{\mu}f_{i}(x,p)= {\cal C}[f_q,f_g](x_i,p_i) \label{eq:RBTeqq}; \\
		& & p^{\mu} \partial_{\mu}f_{HQ}(x,p)= {\cal C}[f_q,f_g,f_{HQ}](x,p) .
        \label{eq:RBTeqQ}
		\label{eq:B_E} 
	\end{eqnarray}
where $f_i(x,p)$ is the phase space one-body distribution function for the $i$-th parton species ($i=q,g$).

We implement only elastic $2\leftrightarrow 2$ collisions:
\begin{align*}
\mathcal C\left[f\right]_{\mathbf{p}} =&
\intop dP_2 \intop dP_{1'} \intop dP_{2'}\left(f_{1'}f_{2'}-f_{1}f_{2}\right) \\ &\times \left|\mathcal{M}\right|^{2}\delta^{\left(4\right)}\left(p_{1}+p_{2}-p_{1'}-p_{2'}\right),
\end{align*}
where $f_i=f(p_i)$ and $dP=d^3p/((2\pi)^3 E_p)$ while
$\mathcal{M}$ denotes the transition amplitude for the elastic processes {$|{\cal M}|^2=16 \pi\,\mathfrak{s}^2 d\sigma/d\mathfrak t$} with {$\mathfrak{s}$} and  {$\mathfrak t$} the Mandelstam variables.

In the collision integral $\mathcal{C} [f_q, f_g](x,p)$ for gluon and light quarks, the total cross section $\sigma_{\rm tot}$ is determined locally in order to keep the ratio $\eta/s=1/(4\pi)$  fixed during the evolution of the QGP (see Refs.~\cite{Plumari:2019gwq,Plumari:2015cfa,Ruggieri:2013ova, Nugara:2025ueb} for more details). 
The analytical relation between $\eta$, the temperature $T(x)$ and $\sigma_{\rm tot}$ is obtained via the Chapman--Enskog approximation at second order for massive particles:
\begin{equation}
\eta(x)=f(z)\frac{T(x)}{\sigma_{\rm tot}(x)};
\label{eq:sigma}
\end{equation}
see Ref. \cite{Plumari:2012ep, Parisi:2025gwq} for details on $f(z)$ with $z(x)=m/T(x)$; in the limit $z\to0$, this function goes to $f(z)\to 1.258$.
In this way, we evolve of a one-component fluid with specified $\eta/s$, simulating a
viscous hydrodynamics evolution at least for sufficiently small $\eta/s$ \cite{Huovinen:2008te,El:2009vj,Plumari:2015sia,Gabbana:2019uqv}; it has also been shown that this approach is able to describe the dynamical evolution in a wide range of $\eta/s$, even below $\eta/s=0.05$ \citep{Gabbana:2019uqv}.
Assuming local equilibrium and being $s=\varepsilon/T +n(1-\ln \Gamma)$, with $s$ entropy density, $\varepsilon$ energy density, $n$ the particle density in the Local Rest Frame and $\Gamma$ the fugacity, Eq.\eqref{eq:sigma} turns into 
\begin{align}
\begin{split}
\sigma = f(z)\frac{T}{[\varepsilon/T +n(1-\ln \Gamma)]\eta/s}.
  \label{eq:etas_sigma_1}
\end{split}
\end{align}

The collision integral for the bulk ${\cal{C}}[f_q, f_g](x,p)$ in the right-hand side of Eq. (1)
discards the impact of charm quarks on the bulk dynamics, which is known to be a quite a solid approximation due to the small number of HQs with respect to the light partons. For the same reason, we discard the scattering between two HQs.\\
When the medium is under local thermal equilibrium, we can define the drag coefficient ${ A}({\bf p},T)$ of HQs, and, 
assuming fluctuation-dissipation theorem (FDT), the diffusion coefficient in momentum space $D_p$ takes the simple form $D_p(p,T) = A(p,T) E(p)T$, where $A(p,T) = {A_i p_i}/{p^2}$.

\begin{figure}
    \centering
    \includegraphics[width=1.0\linewidth]{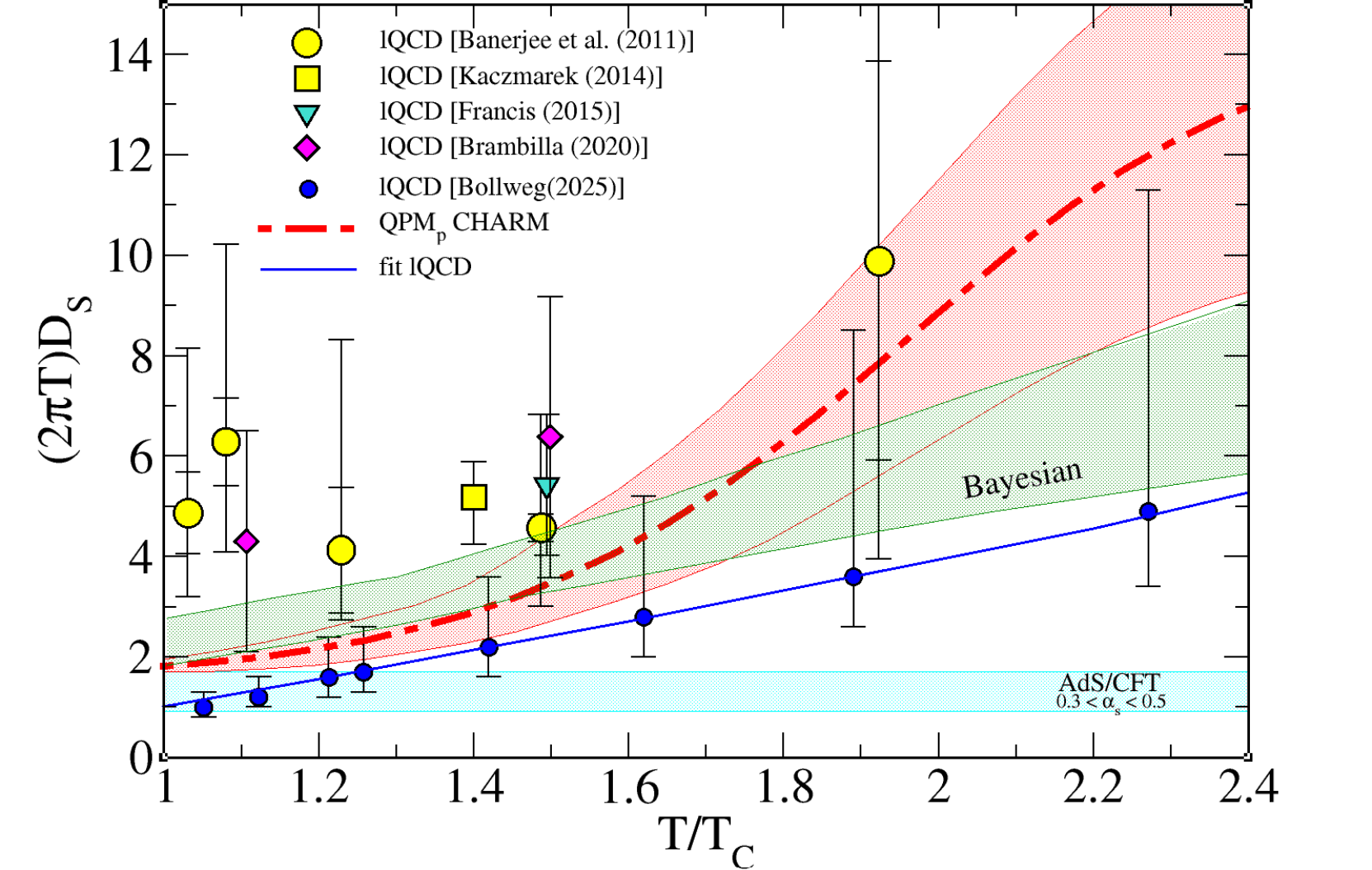}
    \caption{Blue circles are unquenched lQCD data \cite{HotQCD:2025fbd}, blue solid line: fit to lQCD data by a tuned effective coupling $g(T)$ in a Quasi-particle model (see text), referred in the text as $D_s^\text{lQCD}(T)$. Red dot-dashed line and uncertainty band: $2\pi TD_s$ of extended version of Quasi Particle Model (QPMp) \cite{Sambataro:2024mkr, Sambataro:2025obe}.
    Cyan band: AdS/CFT estimate \cite{Gubser:2006qh,Horowitz:2015dta,Casalderrey-Solana:2006fio}. Green band: bayesian analysis from Ref. \cite{Xu:2017obm}. Symbols report the quenched lQCD data taken from Ref.s \cite{Banerjee:2011ra,Kaczmarek:2014jga,Francis:2015daa,Brambilla:2019tpt,Brambilla:2020siz}.}
    \label{fig:2piTDS}
\end{figure}

 In the static limit $p\to 0$, we define $\gamma = A(p\to0)$ and the diffusion coefficient $D_p$ in momentum space is related to a spatial diffusion coefficient $D_s$ via:
\begin{align}
\begin{split}
   D_s ( p \rightarrow 0) = \frac{T^2}{D_p} = \frac{T}{M_{HQ}\gamma} = \frac{T}{M_{HQ}}\tau_{eq}.
    \label{eq:Ds}
\end{split}
\end{align}
In the collision integral of Eq.(\ref{eq:RBTeqQ}) we will consider the scattering matrices at tree level for the processes $g + Q \to g + Q$ and $q(\bar q) + Q \to q(\bar q) + Q$, with finite masses $m_c, m_q, m_g$, which entail a $\mathcal M(s,t,u) \propto g(T)^2$ ~\cite{Sambataro:2024mkr,Sambataro:2025obe} and tune $g(T)$ to reproduce the desired values of $D_s(T)$. 
\\
In this work, we study the charm dynamics in two coupling regimes: $2\pi TD_s=1$, which corresponds to the lower limit of the AdS/CFT estimates (see cyan band in Fig. \ref{fig:2piTDS}) 
\cite{Gubser:2006qh,Horowitz:2015dta,Casalderrey-Solana:2006fio} and the $D_s^\text{lQCD}(T)$ of the recent lQCD data (central values) \cite{Altenkort:2023oms,Altenkort:2023eav,HotQCD:2025fbd} (see blue circles and solid line in Fig. \ref{fig:2piTDS}).

We notice that the upper values of the lQCD calculation, the recent QPMp model \cite{Sambataro:2024mkr}, as well as the Bayesian analysis from phenomenology \cite{Xu:2017obm} hint at larger values of $2\pi T D_s$ and hence of the thermalisation time, but are still able to provide a good agreement to experimental observables $R_{AA}$ and $v_2$ \cite{Sambataro:2025obe};
we will present a more complete study in a following longer paper\cite{Chen:para} including further interaction regimes.

We remind that a bulk plus HQ system is dominated by two different time scales. On the one hand, $\tau_{eq}^\text{bulk}$ characterises the bulk evolution and represents the typical time within which the system approaches the dynamical attractor. In the context of RBT approach, previous works \cite{Nugara:2023eku, Nugara:2024net} showed that the relaxation time defined as the average collision time per particle $\tau_\text{coll}$ is equivalent to the $\tau_{eq}$ introduced in hydro and RTA and defined as $\tau_{eq}\propto (\eta/s)/T$.
On the other hand, the relaxation time of the HQs in a thermal medium is usually defined as the momentum-dependent $\tau_{eq}(p,T) = 1/A(p,T)$. In order to define a unique relaxation time for the HQs, we fix $\tau_{eq}= 1/A(\langle p\rangle, T)$, where $\langle p\rangle$ is the average momentum. Notice that other definitions of $\tau_{eq}$ could be proposed, such as $\tau_{eq}=1/\gamma$ or the HQ mean free time $\tau_{coll}^{HQ}$. However, we have found that all these definitions lead to analogous results for the dynamics toward the attractor that we will discuss in Sect. 3.5. 

From the numerical point of view, the relativistic Boltzmann equation is solved in the RBT framework by sampling the distribution function with a large number ($10^6$--$10^7$) of test particles; the full collision integral is implemented following the stochastic algorithm \cite{Xu:2004mz, Ferini:2008he} on discretised space-time, see Refs: \cite{Nugara:2023eku, Nugara:2024net}. 
In the following, we study the evolution of the charm in the simplified scenario of Bjorken flow, which considers an one-dimensional boost-invariant system evolving along the longitudinal direction; therefore, we adopt Milne coordinates $(\tau,x,y,\eta)$ and consider a box expanding in the longitudinal direction itself. In the transverse plane, we consider a square of $4.2\times4.2$ fm$^2$ , with periodic boundary conditions.
This corresponds to simulate an infinite system in the transverse direction, which is equivalent to a purely 1D expanding system. For the results here shown, we use $\tau_0=0.2$ fm, $\Delta x=\Delta y=0.2$ fm, $\Delta \eta=0.32$; each physical case is obtained running 100 numerical events,  with $ 10^7 \,N_{test}$ particles for the bulk and $N_{test} = 3\times 10^5$ for the heavy flavour sector.

\section{ Dynamical Attractors for charm}


In coordinate space, both the bulk matter and the charm are distributed uniformly in the transverse plane and in a space-time rapidity interval $\eta_s\in [-2.0, 2.0]$. In momentum space, the bulk is initialised as a thermal Boltzmann distribution $f_0(p)\propto \exp( -\sqrt{\vec{p}^2 +m^2}/T_0 )$, with $T_0=0.5$ GeV, resembling typical initial temperatures at LHC energy.

Since we are interested in the possible appearance of universal behaviour and dynamical attractors for charm quarks, we explore a variety of initial conditions in momentum space:\\

$\bullet$ {\bf FONLL}, $f_{\text{FONLL}}(p_T)$ and $Y=\eta_s$; $f_{\text{FONLL}}(p_T)$ is obtained by setting $2.75$ TeV as the beam energy and the PDF set NNPDF30$\_$nlo$\_$as$\_$0118 and taking the central value. This is the typical initial condition used in HQ transport simulations and is able to describe the $D$-meson spectra in proton-proton collisions after fragmentation \cite{Cacciari:2012ny};\\

$\bullet$ {\bf EPOS4HQ}, $f_{\text{EPOS4HQ}}(p_T)$ and $Y=\eta_s$. The $p_T$-distribution is the one in $pp$ collisions of the Monte Carlo simulator EPOS. It includes for HQ productions the Born processes and both space-like and time-like cascades. The $f_{\text{EPOS4HQ}}(p_T)$ is consistent with FONLL for high $p_T$, but predicts a large bump localized in the low $p_T$ ($<2\,\rm GeV$ region), for details Ref. \cite{Zhao:2024ecc}. \\

In order to quantify the deviation with respect to an initial equilibrium distribution, we consider also initial Boltzmann distributions in 2D and 3D, as usually employed in the study of attractors in the light sector \cite{Strickland:2017kux, Nugara:2023eku}:\\

$\bullet$ {\bf 2D-th},  2D transverse Boltzmann distribution $$f(p_T)\propto \exp\left({-\sqrt{p_T^2 +m^2}/T_{0,HQ}}\right)$$ with $T_{0,HQ}=T_{bulk} = 0.5$ GeV and $Y=\eta_s$;\\

$\bullet$ {\bf 3D-th ($\boldsymbol {T_0=0.5}$ GeV)}, 3D Boltzmann distribution  $$f(p)\propto \exp\left({-\sqrt{\vec{p}^2+m^2}/T_{0,HQ}}\right)$$ with $T_{0,HQ}=T_{bulk} = 0.5$  GeV;\\

$\bullet$ {\bf 3D-th ($\boldsymbol {T_0=1.0}$ GeV)}, 3D Boltzmann distribution $$f(p)\propto \exp\left({-\sqrt{\vec{p}^2+m^2}/T_{0,HQ}}\right)$$ with $T_{0,HQ} = 1.0$ GeV, which gives an effective temperature corresponding to a charm average energy per particle similar to the FONLL, but thermally distributed in $\vec p$.

\subsection{Evolution of effective temperature}
As a first investigation about the possible thermalisation of the HQs, we extract an effective temperature $T_{\text{eff}}$ for the HQs and study how it approaches the bulk temperature. We anticipate, that of course, having $T_\text{eff}\approx T_\text{bulk}$ does not imply a full thermalisation, which could be inferred only by studying the whole distribution function or, equivalently, its momentum moments, as discussed in the next Section. We define $T_{\text{eff}}$ according to the standard matching condition starting from the ratio of energy and particle density of the HQ distribution:
\begin{equation}
     \frac{\varepsilon_{HQ}(\tau)}{n_{HQ}(\tau)} =3T_{\text{eff}} +m_{HQ} \frac{K_1(m_{HQ}/T_\text{eff})}{K_2(m_{HQ}/T_\text{eff})}.
\end{equation}

\begin{figure}
    \centering
    \includegraphics[width=1.\linewidth]{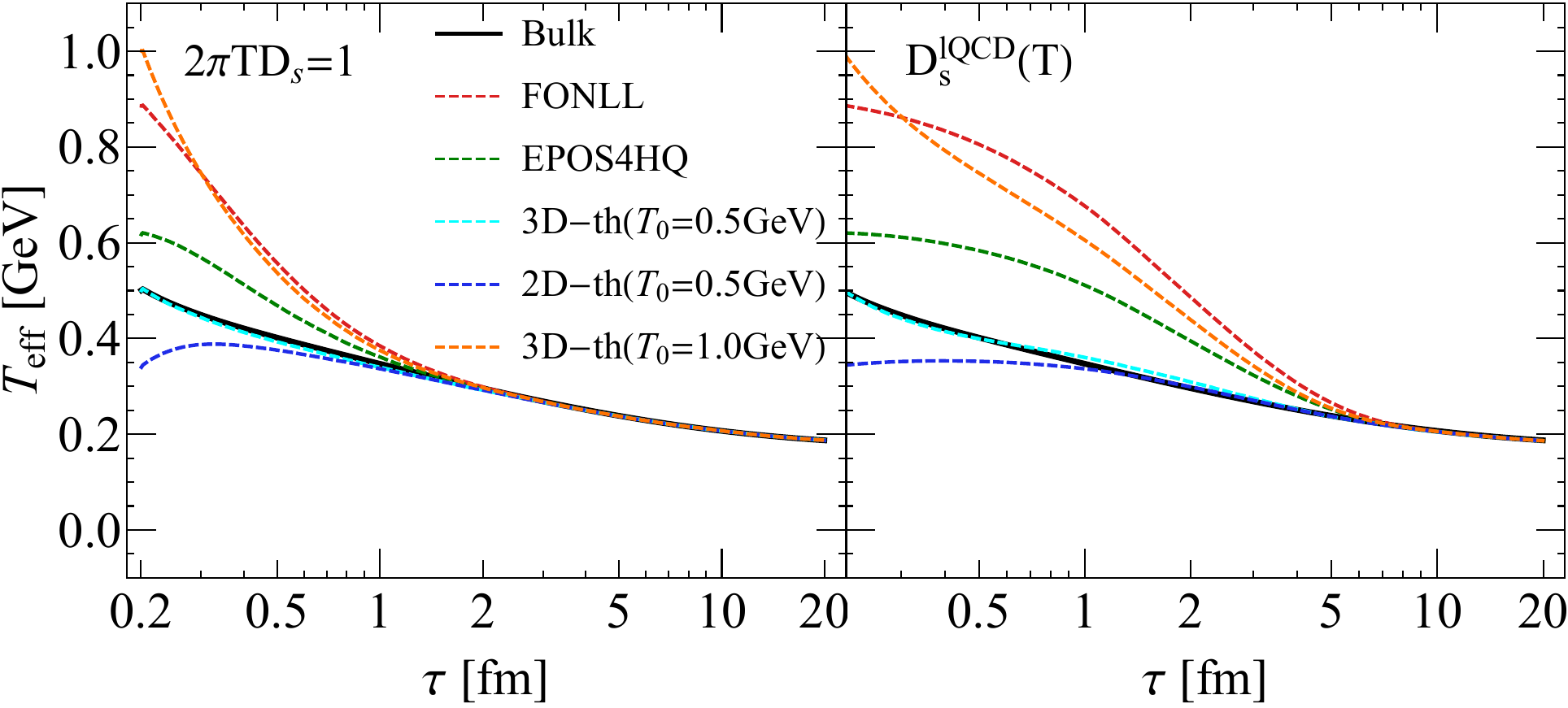}
    \caption{The time evolution of the effective temperature of the charm sector with $2\pi TD_s=1$ (left) and $D^\text{lQCD}_s(T)$ (right) for different initial conditions (coloured dashed lines) compared to the bulk temperature (black solid line).}
    
    \label{fig:fig_2}
\end{figure}

In Figure \ref{fig:fig_2}, we show the $T_\text{eff}$ evolution for the different initial conditions illustrated above and for both the strong coupling regime $2\pi T D_s=1$ (left panel) and $D_s^\text{lQCD}(T)$  (right panel). 
As shown in the left panel, in the strong interaction regime all the different initial conditions, corresponding also to different initial $T_\text{eff}$, reach $T_\text{bulk}$ in 1--1.5 fm, apart from the 3D-th ($T_0=0.5$ GeV) case in which the HQ distribution is initially already in equilibrium with the same temperature of the bulk. Notice also that the FONLL case after less than 0.5 fm has a similar time evolution of the 3D-th case with the high temperature $T_0=1.0$ GeV.  In the right panel, we show the results for the more realistic $D_s^{lQCD}(T)$. This case corresponds to a weaker interaction (at $T>T_c$) and we notice that this leads to a significantly slower equilibration to $T_\text{bulk}$ that occurs in about 5 fm for the FONLL, EPOS4HQ and 3D-th ($T_0=1.0$ GeV), while it is much faster for the 2D-th (that however is not a realistic case for charm). Again, the 3D-th ($T_0=0.5$ GeV) case is nearly identical to the bulk one, 
meaning that if HQ are put in equilibrium with the bulk, the scattering rate is so high that they will be able to keep it during the expansion and cooling of the bulk.

We note that even if new lQCD have a $2\pi TD_s$ at $T\simeq T_c$ very close the AdS/CFT, its $T$ dependence still entails a time scale for charm evolution quite longer than the AdS/CFT case.
Furthermore, we highlight here that, as well known \cite{Nugara:2024net, Ambrus:2022koq}, this 1+1D boost-invariant model is suitable to describe the evolution of a full 3D medium up to $t\approx R$, where $R$ is the typical transverse length scale of the underlying bulk; afterwards, the transverse expansion starts to play a role bringing the system to a weaker interaction regime toward the full decoupling. Our findings suggest that in the AdS/CFT strong coupling regime, HQs seem to equilibrate with the bulk before the decoupling for most physical systems ($R>1.5-2\,\rm fm$); 
instead, in the lQCD case, the more realistic FONLL initial conditions lead to a later effective thermalisation ($\tau \sim 5$ fm), a time scale comparable to the radius of a semi-central Pb–Pb collision, which could indeed approach thermalisation. However, this may no longer hold for smaller systems, such as peripheral Pb–Pb or O–O collisions.

\subsection{Momentum Moments}

Following what has been done for the bulk, we introduce the momentum moments of the distribution function to quantify the approach to equilibrium of $f_{HQ}(p)$ \cite{Strickland:2018ayk} as
\begin{equation}\label{eq:momentum_moments}
\mathcal{M}^{mn}(\tau)=\int dP \, (p\cdot u)^n \, (p\cdot z) ^{2m} \, f_{HQ}(\tau,p),
\end{equation}
with $dP=d^3p/\left((2\pi)^3p^0\right)$; $u^\mu$ is the particle four-flow defined by the Landau matching condition $T^{\mu}_\nu u^\nu=\varepsilon u^\mu$ that, for Bjorken flow in Milne coordinates,  reduces to $u^\mu=(1,0,0,0)$, while $z^\mu=(0,0,0,1)$.
It is useful to define the normalised moments $\overline M^{mn}=\mathcal{M}^{mn}/\mathcal{M}^{mn}_{eq}$ \cite{Strickland:2019hff}, where the moments are rescaled by their corresponding equilibrium values $\mathcal{M}^{nm}_{eq}$:
\begin{equation}\label{eq:momentum_moments_eq}
\mathcal{M}_{eq}^{mn}(\tau)=\int dP \, (p\cdot u)^n \, (p\cdot z) ^{2m} \, f_{eq, HQ}(p; T_\text{eff},\Gamma_\text{eff}),
\end{equation}
where we use the effective temperature and fugacity of the HQs to compute the equilibrium value for the moments.
For a massive Boltzmann distribution with particle number conservation the equilibrium moments are given by {\cite{Strickland:2019hff}:

\begin{align}
\begin{split}
\mathcal M^{nm}_{eq}(\tau)& = \frac{\text{dof}\,\ T^{n+2m+2}}{2\pi^2\ (2m+1)}\Gamma\\
&\times\int dp\ p^{(n+2m+1)}\ \left(1+\frac{\zeta^2}{p^2}\right)^{(n-1)/2} e^{-\sqrt{p^2+\zeta^2}}
  \label{eq:moment_eq}
\end{split}
\end{align}
where $\zeta=m/T$, $\text{dof}$ is the number of degrees of freedom and the $\Gamma $ the fugacity.

In particular, due to the matching conditions:
$$ n = n^\mu u_\mu,\qquad T^{\mu}_\nu u^{\nu} = \varepsilon u^{\mu}, $$
we have that $\overline M^{10}=n/n_{eq}=1$ and $\overline M^{20}=\varepsilon/\varepsilon_{eq}=1$. 
We will focus also on the stress tensor anisotropy $2T_{zz}/(T_{xx}+T_{yy})$, $T_{ij}$ being the $i,j$ component of the energy-momentum tensor, which supplies more direct quantitative information about the degree of isotropisation of the system.

In order to have a more detailed understanding of the thermalisation process, 
we will show two of the first non-trivial moments $\overline M^{21}$ and $\overline M^{22}$. Analogous findings have been obtained for the other moments not shown, except for $\overline M^{n0}$, which, as previously studied in hydro, RTA and RBT \cite{Strickland:2017kux, Nugara:2023eku} for the bulk, do not show far-from-equilibrium convergence, but a simple decay to equilibrium. In Figure \ref{fig:fig_3} different initial conditions for
the charm sector  correspond to dashed lines with different colours, while the black solid line shows the evolution of the moments of the bulk which allows to understand how the HQs dynamics relates to the bulk one.\\
We observe, in Fig. \ref{fig:fig_3} (left panels) that in the strong coupling limit all the cases considered approach a universal behaviour between 1--2 fm. In particular, we see that the two thermal cases with $T_0=0.5$ GeV meet even before ($\tau \sim$ 0.7 fm), while the 3D-th ($T_0=1.0$ GeV), the EPOS4HQ and the FONLL follow a similar time evolution (as seen for the temperature). Notice that the 3D-th ($T=0.5$ GeV) case, whose temperature evolution is identical to that of the bulk, follows now a distinct pattern with respect to the bulk itself. It is far from trivial that higher order moments reach the attractor behaviour in a shorter time with respect to the lower order ones, despite achieving thermalisation later: if one considers 0.8 as a partial thermalisation baseline, in the strong coupling case the anisotropy stress tensor reaches the attractor at $\tau=1.5$ fm;  while $\overline M^{22}$ achieves universality at $\tau \approx 1$ fm and partially equilibrates at $t\approx 3$ fm. 
\\
For the more realistic interaction $D_s^\text{lQCD}(T)$ (right panels of Fig. \ref{fig:fig_3}), the universal behaviour is reached only at $\tau \gtrsim5$ fm, when the HQs have almost the same effective temperature. Moreover, the normalised moments never reach the bulk behaviour before equilibration, suggesting that the HQ dynamics never fully couples to the bulk.\\
In more realistic 3+1D systems, 
as mentioned before, the evolution of the system would reach the phase 
of nearly decoupling at $R<\tau<2R$. This suggests that if we consider small systems ($R\approx 1-3$ fm) in the $D^\text{lQCD}_s(T)$ case, we can envisage some memory of the initial conditions of the HQs, 
remaining significantly far from equilibrium. On the other hand, in larger systems ($R\approx 4-5$ fm, typical of mid-central PbPb or AuAu) HQs could nearly lose any memory of the initial state,
but still do not reach a full degree of equilibration. Instead, the $2\pi T D_s=1$ (AdS/CFT regime) likely induces a very high degree of equilibration similarly to the bulk one.

\begin{figure}
    \centering
    \includegraphics[width=1.0\linewidth]{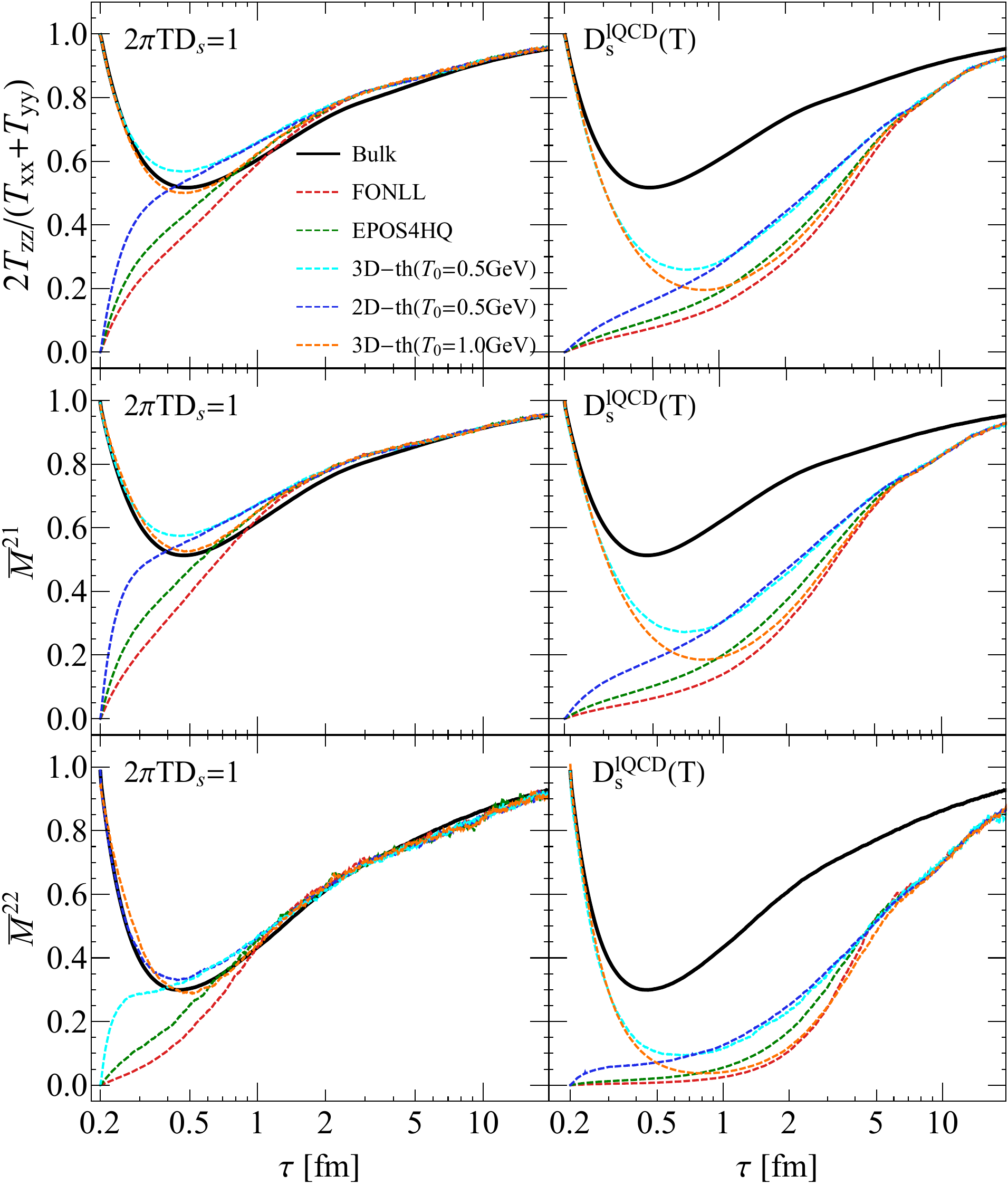}~
    \caption{Left panels: time evolution of moments for $2\pi TD_s=1$; right panels: time evolution of moments for $D_s^\text{lQCD}$. Top panels: The anisotropy of energy momentum tensor $2T_{zz}/(T_{xx}+T_{yy})$. Middle and bottom panels: time evolution of the normalised moments $\overline M^{12}$ and $\overline M^{22}$. Different initial conditions for the charm sector (coloured dash lines) are compared to the bulk one (black solid line).}
    \label{fig:fig_3}
\end{figure}

\subsection{Far-from-equilibrium attractors}
Previous studies in different frameworks in the light sector have shown that the evolution of distinct systems converge to a universal behaviour which does not depend on the initial conditions nor on the interaction regime when studied as a function of $\tau/\tau_{eq}$. At late times, the evolution of physical quantities is known to converge towards the Navier-Stokes limit (late-time attractor), which is mainly related to the fact that the system is already close to equilibrium and thus in the hydrodynamics applicability region. More interestingly, universality can be reached also at early times showing convergence towards the attractor in a region where the system is still far from equilibrium, preventing the application of hydrodynamics: this attractor behaviour is not due to closeness to equilibrium, but is driven by the initial quasi-free longitudinal expansion.\\
In the following, we discuss if the HQ dynamics, which is qualitatively different from the bulk one, exhibits a similar universal behaviour.
Note that heavy quarks in a medium actually do not have a conserved hydrodynamic mode, but their interaction with  the bulk is characterised by different time scales. The relaxation of the energy typically defined by the inverse of the drag coefficient $1/\gamma$ is the fast mode: it characterises the time within which the heavy quark sector approaches the local equilibrium defined by the bulk, as can be seen from the relaxation of the effective temperature. Afterwards, the HQs
momentum is diffused under the expanding bulk according to a different time scale. After this period, when the HQs are totally relaxed to the medium and the medium is also controlled by the long wavelength slow mode, the HQs will follow the bulk evolution. In the following, we are going to study the HQ attractor by means of their own relaxation time $\tau_{eq}^{HQ}=1/A(\langle p\rangle, T)$.

In the previous section, we have studied a broad variety of initial conditions for the HQs, corresponding also to different $T_\text{0,eff}$ and therefore different energy densities.
In Figure \ref{fig:pull_back_attractors}, we show the same set of moments analysed in the previous section for two different initial conditions, the FONLL (solid coloured lines) and the 3D-th ($T_0=0.5$ GeV) (dashed coloured lines), and for a broad range of interaction regimes, going from $2\pi TD_s=1$ to $2\pi TD_s=5$ and including also $D_s^\text{lQCD}(T)$; we did not include all the other initial conditions for the sake of readability.
We have found the existence of an early time attractor, as shown
in Fig. \ref{fig:pull_back_attractors}, which is seen to depend on the initial shape of the distribution function. Indeed, we can see,
for all the studied moments, that for each fixed $f_0(p)$ systems in different interaction regimes converge towards an attractor curve in terms of  the scaled time $\tau/\tau_{eq}$.
However, to our knowledge, this is the first time dynamical attractors are identified in a Brownian component embedded in an evolving medium. In fact, while for the bulk the dynamical attractor is generated by its own interactions, for HQs it is their coupling to the bulk medium that can induce it.
Therefore, this is a case where 
a bulk fluid evolving toward a  dynamical attractors can generate dynamical attractors in the other admixture component coupled to it.

\begin{figure*}
    \centering
    \includegraphics[width=1.0\linewidth]{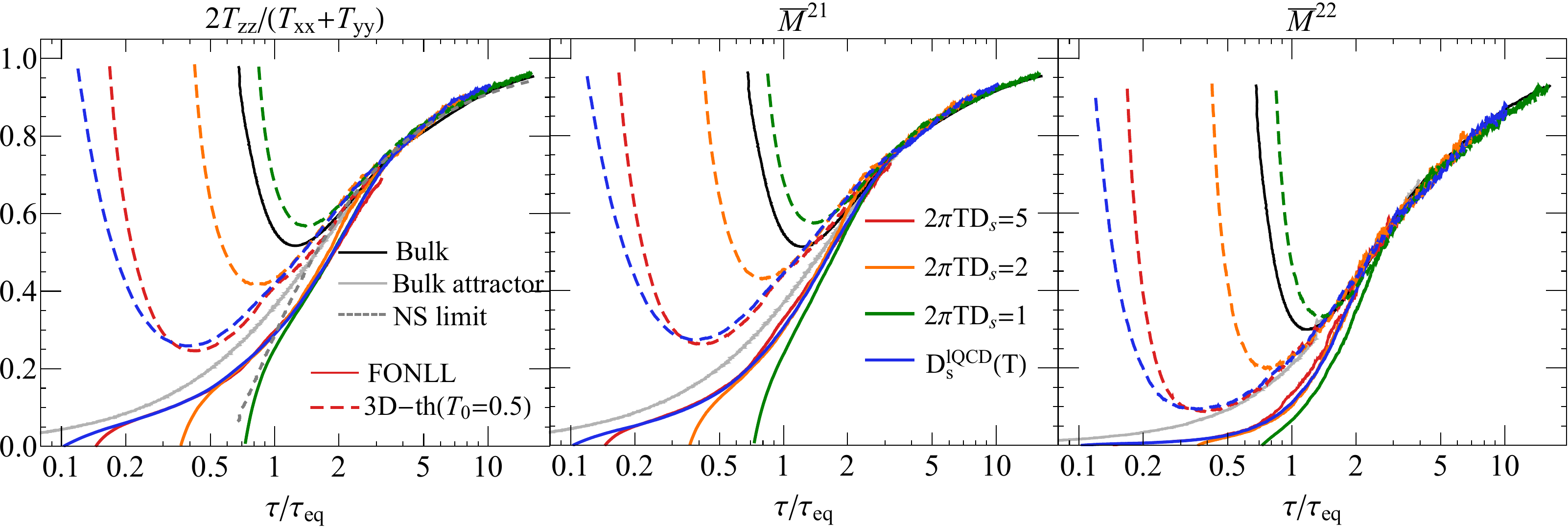}~
    \caption{ Evolution of the stress tensor anisotropy (left panel) and of the normalised moments $\overline M^{21}$ (middle panel) and $\overline M^{22}$ (right panel) of the charm sector in terms of the scaled time $\tau/\tau_{eq}$. Different colours refer to different interaction regimes, from $2\pi T D_s=1$ to $2\pi T D_s=5$, including also $D_s^\text{lQCD}(T)$; dashed lines refer to the 3D-th $T_0=0.5$ GeV initial condition; solid lines to the FONLL initial distribution. The blue solid line report the bulk moments' evolution, while the grey solid line the bulk attractor curve.}
    \label{fig:pull_back_attractors}
\end{figure*}

Finally, we have included in Fig.\ref{fig:pull_back_attractors} our numerical evaluation of the bulk attractor (solid grey line)
checking that it corresponds to the RTA one for a system with $\tau/\tau_{eq} \to 0$ and $P_L(\tau_0)=0$ and $m=0.5$ GeV, according to the prescription of \cite{Alalawi:2022pmg}.
We have also included the Navier Stokes limit for the tensor anisotropy (dashed grey line). We consider the ratio between the shear stress tensor magnitude $\pi=\sqrt{\pi^{\mu\nu}\pi_{\mu\nu}}$, which in the Bjorken flow is the only quantity characterising $\pi^{\mu\nu}$, and the equilibrium pressure $P$:
$$\bar \pi= \frac{\pi}{P} = \frac {4}{3} \frac{\eta}{\tau P}=\frac 4{15} \frac{5\,\hat\gamma(\zeta)\,\eta/s}{\tau\, T \,\hat\gamma(\zeta)}\frac{sT}{P} = \frac{4sT}{15P\hat\gamma(\zeta)} \left({\frac{\tau}{\tau_{eq}}}\right)^{-1} $$
where we have used the non-conformal estimate for the relaxation time $\tau_{eq}= \hat \gamma(\zeta)5(\eta/s)/T$,  where $\hat \gamma(\zeta)=\hat \gamma(m/T)$, as proposed in \cite{Alalawi:2022pmg}. The $\hat\gamma(\zeta)$ function gives the conformal limit $\hat \gamma\to 1$ for $\zeta\to0$; for the temperatures explored in this paper, $1\lesssim\hat\gamma\lesssim1.25$. We have also checked that this relaxation time perfectly agrees with the $\tau_{eq}^\text{bulk}$ as extracted from the code. The stress tensor anisotropy, since the energy-momentum tensor is diagonal, can be expressed as:
\begin{equation}
   \frac{T_{zz}}{T_{xx}+T_{yy}}
   =\frac{1 + \Pi/P - \bar \pi}{1 + \Pi/P +\bar \pi/2}\approx \frac{1- \bar \pi}{1 +\bar \pi/2} 
\end{equation}
where we have used $\Pi/P\ll \pi/P$. We can clearly see that the two distinct pull-back attractors (coloured solid and dashed lines) for the two different initial conditions approach each other only when they reach the Navier-Stokes limit (dashed grey line): this confirms that the observed universality can be considered a late-time convergence. 
We have not included the Navier-Stokes limit for higher order moments since they are beyond the applicability of the theory. 

As anticipated in Section 2, there are two characteristic relaxation times, $\tau_{eq}^\text{bulk}$ and $\tau_{eq}^{HQ}$, which are related to two different transport coefficients, i.e. $\eta/s$ for the bulk and $D_s$ for the HQs: the bulk dynamics is constrained by fixing $\eta/s$, whereas the HQs' one is determined by fixing the $D_s$. This makes the two time scales not directly comparable and in principle their ratio $\tau_{eq}^{HQ}/\tau_{eq}^{bulk}$ depends on both coefficients. However, in our simulations with constant $2\pi T D_s$ we find that the ratio $\tau_{eq}^{HQ}/\tau_{eq}^\text{bulk}$ is approximately constant over the entire evolution, and it is linearly increasing with increasing $2\pi T D_s$. We observe, for the initial conditions and the various coupling regimes explored in this paper, that the late-time behaviour of the bulk (solid grey line) and the HQs attractors (solid and dashed coloured lines) follow the same pattern once we rescale $\tau_{eq}^\text{bulk}\to1.8\,\tau_{eq}^\text{bulk}$.
This rescaling has been considered in Fig. \ref{fig:pull_back_attractors}
and leads to the superposition of the attractor curves for HQ, bulk and Navier-Stokes (dashed grey line) in the late time limit.

\section{Deviation from equilibrium $\delta f/f_{eq}$}
Concerning the dynamics of thermalisation, it is important to quantify the deviation from equilibrium of the full distribution function of the HQs. We show in Fig. \ref{fig:fig_5} the ratio $\delta f_{HQ} (p_T)/f_{HQ}^{eq} (p_T; T_\text{eff})$, where
\begin{equation*}
f_{HQ}^{eq} (p_T; T_\text{eff}) = \text{dof}\int dp_z \exp\left(-\sqrt{p_T^2 + p_z^2 +  m^2}/T_\text{eff}\right),    
\end{equation*}
with $T_\text{eff}$ the HQ effective temperature as defined in section 3.1 and dof is the number of degrees of freedom. A precise determination of this quantity is useful to study whether the condition $\delta f_{HQ}/f_{HQ}\ll 1$, which is one of the basic requirements to guarantee the applicability of hydrodynamics, is fulfilled.
We show the ratio of the different initial distributions explored in this paper ($\tau_0=0.2$ fm) in the top panels. Already at $\tau=0.4$ fm (middle panels)  one can observe relevant deviations due to the evolution and the early particle scatterings, which are obviously less pronounced in the $2\pi TD_s=1$ limit (left panel), but quite sizeable also with $D_s^\text{lQCD}(T)$ (right panel).\\

In the lower panels, we show the results at $\tau=8$ fm which is the typical lifetime of the QGP lifetime in PbPb collisions at LHC. Firstly, the distribution functions depart sensitively from the equilibrium value already at $p_T\sim 2$ GeV, even in the strong coupling limit. 
However, quite interestingly,
irrespectively from the initial distribution function $f_{HQ}(p,\tau_0)$ in the $2\pi TD_s=1$ scenario, the curves show a similar deviation from equilibrium for $p_T\lesssim 3.5$ GeV, resembling what has been previously found for the bulk \cite{Nugara:2023eku}. Instead in the $D_s^\text{lQCD}$ case the universality $\delta f_{HQ}/f_{eq}$ is granted up to $\approx 2 $ GeV, while the deviations 
for charm quarks at higher $p_T$ 
are sensitively different for Boltzmann-like and  FONLL-like tails. Furthermore, the AdS/CFT $2\pi T D_s=1$ implies an interaction strong enough to guarantee $\delta f/f_{eq} \leq 0.25$  up to $p_T\lesssim 2$ GeV; while, if one goes in the $D_s^\text{lQCD}(T)$ case, the deviation is $\approx$10\% already at $p_T\approx $1.5 GeV and rapidly increases becoming as large as 100\%
at $p_T\simeq 3\,\rm GeV$.


To identify according to which power law the $\delta f$ increases, we have fitted our numerical results with:
\begin{equation}
\delta f_{HQ} (p_T)/f_{eq}= x_0+\alpha p_T^{\beta}    
\end{equation}
We find, see Table \ref{tab:power_law_fit}, that in the $2\pi TD_s =1$ case as well as for the Boltzmann-like initial conditions with $D_s^\text{lQCD}(T)$ the power is approximately quadratic.
This suggests that in developing the correction to the equilibrium distribution function, the expansion should be extended at least up to the second order. 
Moreover, for $D_s^\text{lQCD}(T)$ the power $\beta$ becomes as high as $\beta > 4$ in the case of a realistic initial FONLL distribution function; a very strong deviation from equilibrium already at moderate $p_T$ that casts doubts on the applicability of viscous hydrodynamics to charm.


\begin{table}
    \centering
    \begin{tabular}{c|ccc}
         & $x_0$ & $\alpha$ [GeV$^{-1}$]  & $\beta$ [GeV$^{-2}$] \\
        \toprule
        $2\pi TD_s=1$ & -0.028 & 0.027 & 2.2 $\pm$ 0.2 \\
        \midrule
        $D_s^\text{lQCD}(T)$Boltz-(0.5 GeV) & -0.046 & 0.049 & 2.4 $\pm$ 0.1  \\
        \midrule
        $D_s^\text{lQCD}(T)$-FONLL & -0.011 & 0.007 & 4.6 $\pm$ 0.2  \\
    \end{tabular}
    \caption{Fit parameters to $\delta f_{HQ}/f_{eq}$ by $x_0+\alpha p_T^{\beta}$ for different values of $D_s(T)$ and initial charm distribution functions at $\tau= 8\, \rm fm$.}
    \label{tab:power_law_fit}
\end{table}



\begin{figure}
    \centering
    \includegraphics[width=1.0\linewidth]{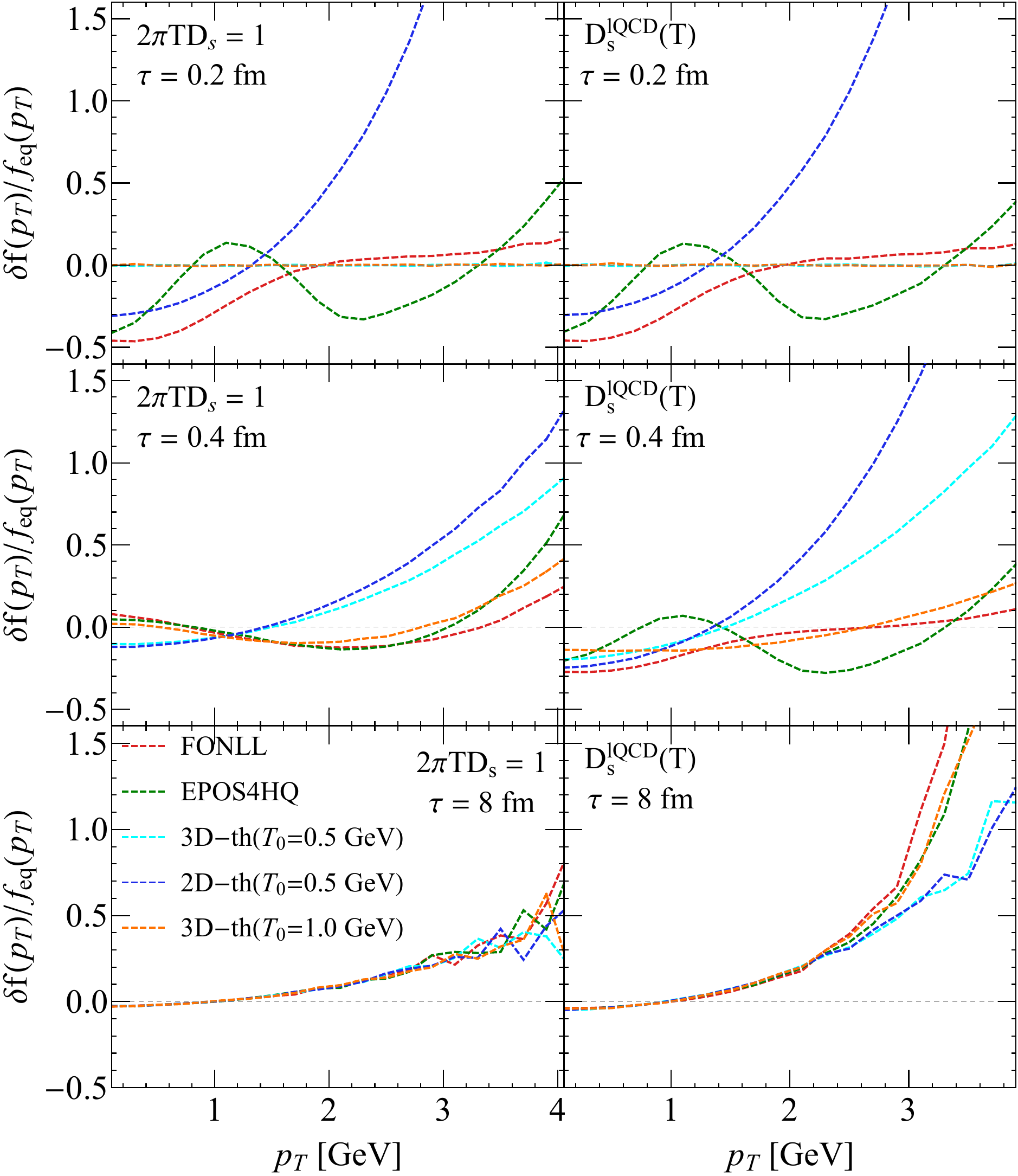}
    \caption{$\delta f(p_T) /f_{eq}(p_T)$ for different initial charm distribution functions (FONLL, EPOS4HQ, Boltzmann) for $2\pi TD_s=1$ (left panels) and $D_s^\text{lQCD}(T)$ (right panels) at different time from top ($\tau=0.2$ fm), middle ($\tau=0.4$ fm, i.e. after the initial strong longitudinal expansion) to bottom ($\tau= 8\,\rm fm)$.   }
    \label{fig:fig_5}
\end{figure}

\section{Conclusions and Outlook}
In this Letter, we have extended for the first time the study on attractors to the heavy quark sector, in order to investigate the hydrodynamisation/thermalisation of HQs and the possible emergence of universal behaviour. This could help to understand the similarities found between the light and the heavy flavour observables, such as in the anisotropic flows $v_n$. In the context of the Relativistic Boltzmann Transport approach, we have studied the timescales within which the heavy flavour quarks equilibrate to the bulk dynamics, looking at the effective temperature $T_{eff}$ and at several moments $\mathcal{M}^{mn}$ of the HQ distribution  function. We have focused on two most relevant cases: $2\pi TD_s=1$, motivated by AdS/CFT calculations and corresponding to the strongest coupling scenario, and $D_s^\text{lQCD}(T)$, which follows the recent lQCD data that show at $T\approx T_c$ a behaviour similar to AdS/CFT, while approaching pQCD results at larger $T$. 
The first novel result is that HQs exhibit a dynamical attractor and a first non-trivial finding has been that, even if $ 2\pi T D_s^\text{lQCD}(T_c)\simeq 1$ close to AdS/CFT limit, the $T$-dependence found in lQCD is significantly 
large to imply a quite different dynamical evolution of charm towards equilibrium entailing about 3--4 times larger time scale for reaching the dynamical attractors.
In the AdS/CFT case, we observe that the system reaches thermalisation and isotropisation within 1--1.5 fm which corresponds, at least for lower order moments, to the timescale in which the system loses memory about the initial conditions. 
This is likely to have important consequences for the charm quark dynamics as a function of system size of the colliding ions. Such a short time-scale would allow the equilibration of heavy quarks in most collision systems: the Bjorken flow describes the dynamics quite well up to $t\approx R$, which means that for collision systems with $R>2$ fm the partial thermalisaton of HQs would be achieved before the onset of transverse expansion. Quite differently, in the more realistic case of $D_s^\text{lQCD}(T)$ the time scales within which the system partially equilibrates are much longer. Normalised moments reach 0.8 only at $\tau\sim 8$ fm, which is larger than $R$ even for Pb-Pb collisions and would imply to allow for a partial thermalisation only for the largest systems (central collisions of heavy nuclei) and not for light systems.
Therefore, the increased time scale for equilibration with $D_s^\text{lQCD}(T)$ suggests that for light-ion system, such as $^{16}\text{O}$-$^{16}\text{O}$, there could be significant non-equilibrium even in the low $p_T$ region; however a solid statement in this direction requires a dedicated study in 3+1D.
\\
We have looked also at the HQ dynamics in terms of $\tau/\tau_{eq}$ and found that, for fixed initial conditions, a far-from-equilibrium attractor is present if one changes the interaction regime. Quite interestingly, the far-from-equilibrium attractors are different if one starts with a Boltzmann-like or a FONLL initial distribution in momentum space, with the two curves converging only at late time, when they reach also the bulk attractor and the Navier-Stokes limit.\\
Finally, we have looked at the deviation of the HQ distribution function $\delta f_{HQ}/f_{eq}$ at late time $\tau=8$ fm, finding that in 1+1D equilibration is achieved for $p_T\lesssim 2$ GeV in the strong coupling limit and for $p_T\lesssim 1.5$ GeV in the lQCD case. However, we find that at intermediate $p_T$ the $\delta f_{HQ}/f_{eq}$ is always positive and increases at least with a quadratic power for the AdS/CFT case, but again the $D_s^\text{lQCD}(T)$ entails a quite larger correction with a power $\beta \simeq 4.5$ which means a $\delta f_{HQ}/f_{eq}$ of about a 100\% already at $p_T\simeq 3 \,\rm GeV$, seriously challenging the applicability of hydrodynamics for the charm sector in this realistic case. 

Recently, a significant breakthrough has been achieved within the AdS/CFT approach deriving a self-consistent equation for HQ dynamics, named Kolmogorov equation \cite{Rajagopal:2025ukd}, that including non-Gaussian fluctuations ensure a dynamics toward equilibrium in an AdS/CFT framework and more generally in any quantum field theory \cite{Rajagopal:2025rxr}. It would be very relevant to see if such an approach confirm the main findings presented in this Letter, especially for $2\pi T D_s=1$.
However, a solid assessment of the existence of attractors and a quantitative determination of the deviation from equilibrium of HQ quarks requires a study in 3+1D that is currently in progress.

\section{Acknowledgments}
We acknowledge the funding from UniCT under PIACERI Linea d'intervento 1 (project M@RHIC).
We thank the Galileo Galilei Institute for Theoretical Physics for the hospitality and the INFN for partial support under SIM project.
S.C. thanks Dr. Jiaxing Zhao for EPOS4 Data. 
\bibliography{ref}

\end{document}